\pdfoutput=1

\documentclass
[reprint,aip,groupedaddress,citeautoscript]{revtex4-1}

\usepackage[colorlinks=true,bookmarks=false,
 pdfauthor={I. Muneta, H. Terada, S. Ohya, and M. Tanaka},
 pdftitle={Anomalous Fermi level behavior in GaMnAs at the onset of ferromagnetism}
 ]{hyperref}
\usepackage{graphicx}
\usepackage[T1]{fontenc}
\usepackage{mathptmx}

\begin{document}

\title{Anomalous Fermi level behavior in GaMnAs at the onset of ferromagnetism}

\author{Iriya Muneta}
\email{muneta@cryst.t.u-tokyo.ac.jp}
\affiliation{Department of Electrical Engineering and Information Systems,
The University of Tokyo, 7-3-1 Hongo, Bunkyo-ku, Tokyo 113-8656, Japan}

\author{Hiroshi Terada}
\affiliation{Department of Electrical Engineering and Information Systems,
The University of Tokyo, 7-3-1 Hongo, Bunkyo-ku, Tokyo 113-8656, Japan}

\author{Shinobu Ohya}
\affiliation{Department of Electrical Engineering and Information Systems,
The University of Tokyo, 7-3-1 Hongo, Bunkyo-ku, Tokyo 113-8656, Japan}

\author{Masaaki Tanaka}
\email{masaaki@ee.t.u-tokyo.ac.jp}
\affiliation{Department of Electrical Engineering and Information Systems,
The University of Tokyo, 7-3-1 Hongo, Bunkyo-ku, Tokyo 113-8656, Japan}

\begin{abstract}
We present the systematic study of the resonant tunneling spectroscopy on a
series of ferromagnetic-semiconductor Ga$_{1-x}$Mn$_x$As with the Mn content $x$
from $\sim$0.01 to 3.2\%.
The Fermi level of Ga$_{1-x}$Mn$_x$As exists in the band gap in the whole $x$
region.
The Fermi level is closest to the valence band (VB) at $x$=1.0\% corresponding
to the onset of ferromagnetism near the metal-insulator transition (MIT), but it
moves away from the VB as $x$ increases or decreases from 1.0\%.
This anomalous behavior of the Fermi level indicates that the ferromagnetism and
MIT emerge in the Mn-derived impurity band.
\end{abstract}

\date{}

\maketitle

The origin of the ferromagnetism and the metal-insulator transition (MIT) has
been a long-debated issue in the prototype ferromagnetic semiconductor GaMnAs
\cite{PhysRevB.57.R2037, PhysRevB.76.125206,
Richardella05022010, PhysRevB.84.081203}.
Previously, the valence band (VB) conduction picture has been widely accepted
in this material \cite{Dietl11022000},
where the MIT of GaMnAs was understood by the Fermi level crossing over the VB
\cite{PhysRevB.76.125206,PhysRevLett.105.227201}
similarly to p-type GaAs doped with non-magnetic acceptors such as Be or Zn.
The ferromagnetism in GaMnAs has been thought to be
induced by the VB holes interacting with the localized $d$ electrons of the Mn
atoms \cite{Dietl11022000,PhysRevLett.105.227202}.
However, recently, many experiments have shown the strong evidence
that the Fermi level exists in the impurity band (IB) in the band gap
\cite{PhysRevB.84.081203,PhysRevB.65.193312,PhysRevLett.94.137401,
PhysRevB.80.041202, PhysRevLett.97.087208,
PhysRevLett.100.067204,PhysRevLett.102.247202,
PhysRevB.75.155328,PhysRevLett.104.167204,NatPhys.7.342,PhysRevB.76.161201,
PhysRevB.84.121202,PhysRevB.78.075201,PhysRevB.81.045205,Dobrowolska2012},
which requires reconsideration on the above scenario.
Therefore, to clarify the origin of the ferromagnetism and MIT in GaMnAs, it is
essential to precisely investigate the Fermi level position and the VB structure
of GaMnAs in the low Mn content region including the onset of ferromagnetism and
MIT.

Resonant tunneling spectroscopy is a powerful method to investigate the VB
structure and the Fermi level position in GaMnAs with a precision of several
meV.
The advantage of this method is that we can detect energy bands only with the
same wave-function symmetry as that of the p-wave functions of the VB holes.
Also, it is sensitive to the effective mass of the energy bands.
Furthermore, by carefully analyzing the quantum-well (QW) thickness dependence
of the resonant levels, we can avoid the unwanted effects induced at the
surface, which prevent the precise determination of the VB or the Fermi level
position \cite{PhysRevB.64.125304,Richardella05022010}.
Recently, from the resonant tunneling experiments in the double-barrier (DB) QW
heterostructures \cite{PhysRevB.75.155328,PhysRevLett.104.167204} and the
single-barrier structures with an ultra-thin surface GaMnAs layer
\cite{NatPhys.7.342}, it was shown that the Fermi level exists in the band gap in
Ga$_{1-x}$Mn$_x$As with the Mn content $x$ higher than $\sim$5\%.
In this Letter, we carefully analyze the VB structure and the Fermi level
position in a series of Ga$_{1-x}$Mn$_x$As from the unexplored insulating region
($x$=$\sim$0.01\%) to the metallic region ($x$=3.2\%).
We find that the VB structure is not largely affected by the Mn doping and that
the Fermi level never crosses over the VB near the MIT: The Fermi level becomes
closest to the VB at $x$=1.0\% at the onset of the ferromagnetism, but it moves
away from the VB as $x$ increases or decreases from 1.0\%.
This anomalous behavior of the Fermi level is completely different from that of
GaAs doped with other shallow acceptors.

\begin{figure}
\includegraphics{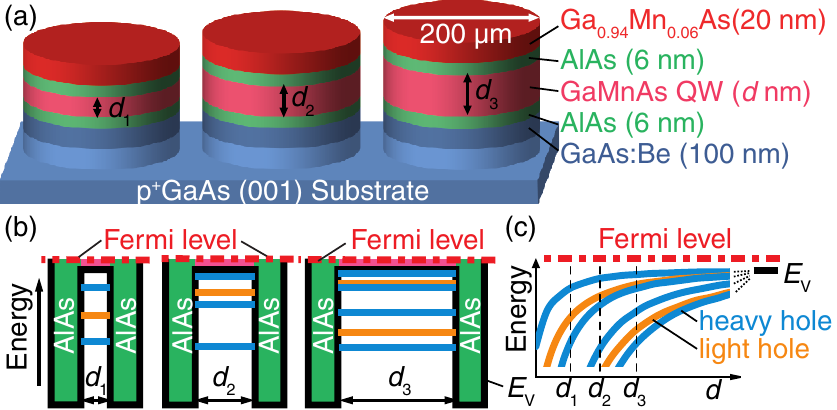}
\caption{\label{fig:1}
(a) Schematic device structures investigated in this study.
$d_1$, $d_2$, and $d_3$ are the GaMnAs QW thickness.
(b) VB lineups of AlAs / GaMnAs QW / AlAs when the QW thickness is $d_1$, $d_2$,
and $d_3.$
The black solid and red dash-dotted lines correspond to the VB top and Fermi
level.
The blue and orange lines correspond to the quantum levels in heavy and light
hole bands, respectively.
(c) $d$ dependence of the quantum levels, where the blue and orange curves
correspond to the heavy and light hole bands, respectively.}
\end{figure}

Figure \ref{fig:1}(a) shows the schematic device structures investigated in this
study.
We grew DB-QW heterostructures composed of (from the top to the bottom)
Ga$_{0.94}$Mn$_{0.06}$As (20~nm) / AlAs (6~nm) / Ga$_{1-x}$Mn$_x$As QW ($d$ nm)
/ AlAs (6~nm) / GaAs:Be (100~nm, Be concentration: 1$\times$18~cm$^{-3}$) by
low-temperature molecular-beam epitaxy (MBE).
Table \ref{tab:1} shows the characteristics of the GaMnAs QW samples examined in
this study.
The Mn content $x$ of sample D-H, F2, and G2 was determined by the x-ray
diffraction measurements for the 100 nm-thick reference GaMnAs samples grown in
the same condition as that for the GaMnAs QW in each DB-QW sample.
We estimated $x$ of sample A-C and C2 from an Arrhenius plot of the Mn flux
\textit{vs.} the K-cell temperature obtained from the flux data of sample D-H,
F2, and G2.
In this study, $x$ was varied from $\sim$0.01\% to 3.2\%.
In this Mn content range, as $x$ increases,
GaMnAs changes from insulating paramagnetic to insulating ferromagnetic at
$x$=$\sim$1\%.
At ~2\%, it starts to show the metallic behavior, in which the resistivity
decreases as temperature $T$ decreases at $T < T_{\rm C}$ (Curie temperature).
The $T_{\rm C}$ values of the Ga$_{1-x}$Mn$_x$As QW in sample D-H, F2, and G2
were obtained from the temperature dependence of tunneling magnetoresistance
(TMR) observed in these devices.
As shown in Table \ref{tab:1}, the Ga$_{1-x}$Mn$_x$As QW with $x$=1-3\% has
lower $T_{\rm C}$ than that of the surface Ga$_{0.94}$Mn$_{0.06}$As (60-70~K)
due to the low $x$ in the QWs.
To obtain the resonant tunneling diodes with various GaMnAs QW thickness $d$ on
a same wafer, we linearly moved a shutter in front of the substrate while
growing the Ga$_{1-x}$Mn$_x$As QW, in which $d$ was varied from 10~nm to 16~nm
within the wafer of $15\times 10$ mm$^2$
\cite{PhysRevLett.87.026602}.
Also, we prepared sample C2 ($x$=$\sim$0.3\%), F2 ($x$=1.6\%), and G2
($x$=2.3\%) with $d$ from 4~nm to 10~nm, where the growth conditions of sample
C2, F2, and G2 are the same as those of sample C ($x$=$\sim$0.3\%), F
($x$=1.6\%), and G ($x$=2.3\%) except for $d$, respectively.
The values of $d$ were estimated from the speed of the shutter.
Although simple structures (GaMnAs / AlAs / GaAs:Be single-barrier
heterostructures) were used in Ref. \onlinecite{NatPhys.7.342}, they are not
applicable for this study due to the low Mn contents $<$ $\sim$2\% in the GaMnAs
QW that we are using here.
The low Mn contents in the surface GaMnAs induce a large depletion thickness at
the surface, making an electrical contact to the GaMnAs layer very difficult.
Thus, the DB-QW structures are the best structure for our current purpose.
The details about the preparation of the samples are described in the
supplemental material \cite{suppl}.

\begingroup
\squeezetable
\begin{table}
\centering
\caption{\label{tab:1}
Parameters and characteristics of the DB-QW samples investigated in this study.
$x$ is the Mn content of the Ga$_{1-x}$Mn$_x$As QW.
$T_{\rm S}$ is the growth temperature of the top AlAs barrier and the GaMnAs QW.
$d$ and $T_{\rm C}$ are the thickness and Curie temperature of the GaMnAs QW,
respectively.
$E_{\rm F}$ is the estimated energy distance between the Fermi level and the VB
top at the $\Gamma$ point in the GaMnAs QW.}
\vspace{1em}
\begin{tabular}{lrcrrcccc}
  \toprule
    Sample.  & $x$ (\%)    & $T_{\rm S}$ (${}^\circ$C) & $d$ (nm) & $T_{\rm C}$
    (K) & $E_F$ (meV) \\ \hline
    Sample A & $\sim$0.01 & 400 & 10 - 16 & paramagnetic & 60 \\
    Sample B & $\sim$0.1  & 350 & 10 - 16 & paramagnetic & 35 \\
    Sample C & $\sim$0.3  & 330 & 10 - 16 & paramagnetic & 25 \\
    Sample C2& $\sim$0.3  & 330 &  4 - 10 & paramagnetic & 25 \\
    Sample D & 1.0        & 265 & 10 - 16 & 25 & 4 \\
    Sample E & 1.2        & 260 & 10 - 16 & 30 & 6 \\
    Sample F & 1.6        & 250 & 10 - 16 & 40 & 10 \\
    Sample F2& 1.6        & 250 &  4 - 10 & 25 & 7 \\
    Sample G & 2.3        & 240 & 10 - 16 & 45 & 17 \\
    Sample G2& 2.3        & 240 &  4 - 10 & 30 & 12 \\
    Sample H & 3.2        & 230 & 10 - 16 & 60 & 25 \\
    \botrule
\end{tabular}
\end{table}
\endgroup

Figure \ref{fig:1}(b) shows the schematic VB profiles of the DB-QW devices shown
in Fig.~\ref{fig:1}(a).
The black solid and red dash-dotted lines indicate the VB top and Fermi level,
respectively.
The blue and orange lines are the resonant levels of the heavy hole (HH) and
light hole (LH), respectively.
Here, the small exchange splitting is neglected for simplicity.
Figure~\ref{fig:1}(c) shows the schematic $d$ dependence of the resonant levels.
The resonant levels converge on the VB top energy $E_{\rm V}$ of the bulk GaMnAs
as $d$ increases.
Since the Fermi level corresponds to the zero bias condition, the Fermi level
position can be determined by measuring bias voltage corresponding to $E_{\rm
V}$ in sufficiently wide $d$.
The bias polarity is defined by the voltage of the top GaMnAs electrode with
respect to the substrate.

\begin{figure}
\includegraphics{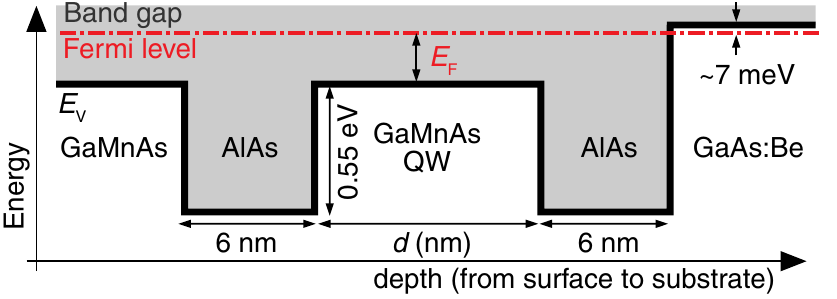}
\caption{\label{fig:2}
VB diagram of the DB-QW devices used in the theoretical calculation.
The black solid and red dash-dotted lines are $E_{\rm V}$ and the Fermi level,
respectively.
The gray region is the band gap.}
\end{figure}

Figure \ref{fig:2} shows the VB diagram of the investigated DB-QW devices
assumed in the following theoretical calculation of the resonant levels.
The black solid and red dash-dotted lines are $E_{\rm V}$ and the Fermi level,
respectively.
The gray region is the band gap.
For the quantum level calculation, we used the $4 \times 4$ Luttinger-Kohn $\bf
k \cdot p$ Hamiltonian \cite{PhysRevB.39.12802} and transfer matrix method
\cite{tsu:562}.
The VB offset between AlAs and GaMnAs was assumed to be 0.55~eV.
We assumed that the in-plain wave vector $\bf k_{\parallel}$ is $\bf 0$ during
the tunneling, because holes are injected from the GaAs:Be electrode in the
negative bias region and holes in the GaAs:Be electrode exist only around the
$\Gamma$ point.
For simplicity, the small band bending was neglected \cite{suppl}.

\begin{figure*}
\includegraphics{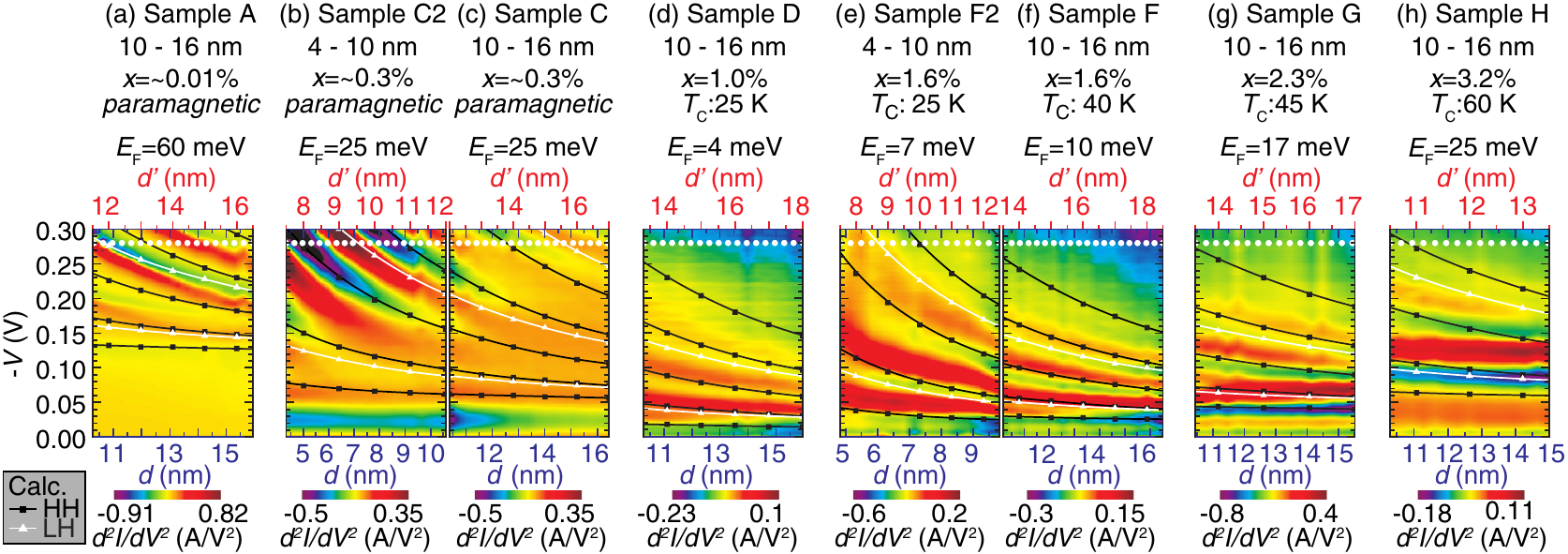}
\caption{\label{fig:3}
(a)-(h) The comparison between the calculated resonant levels and the
experimentally obtained $d^2I/dV^2$ data of (a) sample A, (b) sample C2, (c)
sample C, (d) sample D, (e) sample F2, (f) sample F, (g) sample G, and (h)
sample H as functions of $-V$ and $d$.
The $d^2I/dV^2$ intensity is expressed by color.
Here, these color intensities are extrapolated from the measured data with $d$
corresponding to the white dots shown at the top of these figures.
The connected black and white dots are the calculated resonant peak bias
voltages $V_{\rm R}$ of the HH and LH bands, respectively.
$E_{\rm F}$ and $d^\prime$ are defined in the main text.}
\end{figure*}

Figure \ref{fig:3}(a)-(h) show the color contour maps of the $d^2I/dV^2$-$V$ as
a function of $d$ in sample A, C2, C, D, F2, F, G, and H measured at 3.5~K in
the negative bias region, respectively.
Here, the color-coded intensities are extrapolated from the measured data at the
$d$ values corresponding to the white dots shown at the top of these figures.
The black square and white triangle dots are calculated resonant peak bias
voltages $V_{\rm R}$ of the HH and LH bands, respectively.
$E_{\rm F}$ is the energy distance between $E_{\rm V}$ and the Fermi level.
Clear oscillations due to the resonant tunneling are observed in all the
$d^2I/dV^2$-$V$ curves.
Moreover, the resonant peak bias voltages become smaller as $d$ increases, which
is well reproduced by the calculated $d$ dependence of the resonant peak bias
voltages.
The $E_{\rm F}$ value is estimated by the following procedure.
As $d$ increases to infinity, these resonant levels converge on a certain
voltage $V_{\rm VB}$ corresponding to the valence band edge.
The Fermi level position corresponds to the zero bias.
Experimentally, the measured voltage $V$ is proportional to the energy $E$
(relative to the Fermi level); $V$=$sE$.
Therefore, $V_{\rm VB}/s$ corresponds to the energy distance $E_{\rm F}$ between
the Fermi level and $E_{\rm V}$.
Then, the $d$ dependence of $V_{\rm R}$ is calculated from the equation $V_{\rm
R}$=$sE_{\rm R}$ in individual $d$.
Here, $E_{\rm R}$ is quantum level energy with respect to the Fermi level.
In the fitting, we slightly shifted the $d$ value to $d^\prime$ in order to
obtain the perfect fit.
The value of $d^\prime$ used in the calculation was shown as the upper red ruler
of the contour maps.
The necessity of this shift of $d$ is due to the difficulty in accurately
controlling the relative positions between the substrate shutter and the sample
wafer inside the MBE chamber.
We note that, however, the thickness difference between $d$ and $d^\prime$ is
only less than $\sim$3~nm.
The $s$ value corresponds to the ratio of the total bias voltage to the voltage
applied at the bottom AlAs barrier.
The $s$ values are assumed to be 2.0 in this study.
Considering that $s$ is 2 in ideal DB tunnel junctions \cite{tsu:562} and that
the interstitial Mn and its distribution variation in the QW are sufficiently
suppressed in the low $x$ with the relatively high growth temperature, the
$s$=2.0 assumed in this study is sufficiently reasonable.
The color contour maps of the $d^2I/dV^2$ in all the samples are shown in the
supplemental material \cite{suppl}.

In Fig. \ref{fig:4}, the blue circles are the $E_{\rm F}$ values obtained in
this study.
The green squares are the ones reported in Ref. \onlinecite{NatPhys.7.342}.
The black pentagons are the thermal-activation energy $E_{\rm a}$ of the Mn
acceptors in GaAs obtained by the magneto-transport measurements in Ref.
\onlinecite{blakemore:3352}.
The pink curve is the calculated $E_{\rm a}$ values obtained from the equation
$0.11[1-(x/0.8)^{1/3}]$ mentioned in Ref. \onlinecite{blakemore:3352}.
Here, we selected 0.8\% as the intercept of $x$ to fit the curve to the
experimental data.
The gray dash-dotted, dashed, and solid curves are the calculated $E_{\rm F}$ by
the valence-band anti-crossing model (VBAC) \cite{PhysRevB.75.045203} and the
free-electron approximation.
In the VBAC, the impurity level $E_{\rm Imp}$ are assumed to be 0.1, 0.05, and
0.01 eV, respectively.
Also, the anti-crossing coupling constant $C_{\rm Mn}$ is assumed to be 0.18 eV.
The free-electron approximation calculation for the Fermi level position in the
IB is done by roughly assuming the hole concentration $p$ to be $x$/2 and the
effective mass to be $10m_{\rm e}$ \cite{PhysRevLett.97.087208}, where $m_{\rm
e}$ is the electron mass.
Figure~\ref{fig:4}(b)-(d) show the VB and IB diagrams expected from this study
in the (b) insulating paramagnetic ($x<1\%$), (c) insulating ferromagnetic
($x$=1-2\%), and (d) metallic ferromagnetic ($x>2\%$) regions, respectively.
The black solid curves correspond to the VB.
The blue dotted lines correspond to the upper and bottom edges of the IB.
The blue region represents the IB region.
The red dash-dotted lines correspond to the Fermi level.
In the paramagnetic GaMnAs ($x<1\%$), $E_{\rm F}$ decreases as $x$ increases,
which is quantitatively consistent with the activation-energy-lowering effect
observed in the magneto-transport measurements \cite{blakemore:3352}.
This behavior is the same as the one in the insulating region of the
non-magnetic accepter-doped $p$-type GaAs \cite{hill:1815}.
The Fermi level behavior in the paramagnetic region is caused by the screening
effect due to the heavy Mn doping, which makes the IB position close to the VB.
On the other hand, $E_{\rm F}$ increases as $x$ increases in the
ferromagnetic GaMnAs with $x > 1\%$, which means that the Fermi level moves away
from the VB.
This behavior is qualitatively explained by the VBAC model [the gray solid curve in
Fig.~\ref{fig:4}(a)].
The quantitative discrepancy between the experimental and calculated Fermi level is probably because
is probably because this calculation does not take into account the screening
and many body effects.
This Fermi level behavior in the ferromagnetic region means that the
IB truly exists in the ferromagnetic GaMnAs and the anti-crossing interaction
is induced by the electron-electron interaction \cite{PhysRevB.78.075201}.

\begin{figure}
\includegraphics{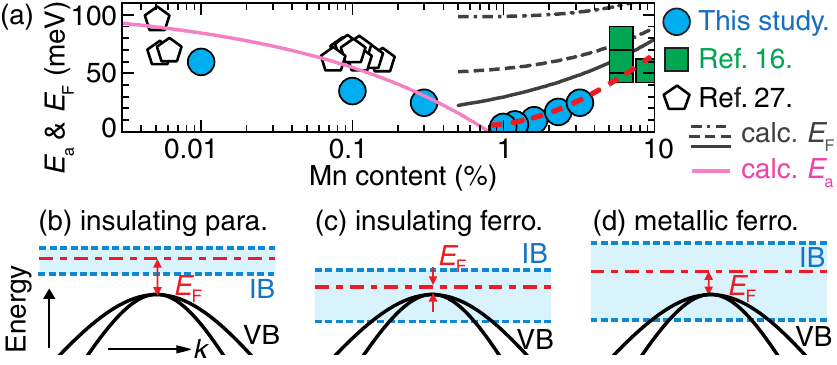}
\caption{\label{fig:4}
(a) The blue circles and green squares are the $E_{\rm F}$ values obtained in
this study and in Ref. \onlinecite{NatPhys.7.342}, respectively.
The black pentagons are the $E_{\rm a}$ values reported in Ref.
\onlinecite{blakemore:3352}.
The pink curve is the calculated $E_{\rm a}$ obtained from the equation
$0.11[1-(x/0.8)^{1/3}]$.
The gray dash-dotted, dashed, and solid curves are the calculated $E_{\rm F}$
with respect to $E_{\rm V}$ by the VBAC model \cite{PhysRevB.75.045203} and the
free-electron approximation.
In the VBAC, $E_{\rm Imp}$ are assumed to be 0.1, 0.05, and 0.01 eV, respectively.
The red dashed curve connects the $E_{\rm F}$ values in $x > 1\%$.
(b)-(d) The VB and IB diagrams of GaMnAs derived from this study in the (b)
insulating paramagnetic ($x<1\%$), (c) insulating ferromagnetic ($x$=1-2\%), and
(d) metallic ferromagnetic ($x>2\%$) regions.
The black solid curves are the VB.
The blue dotted lines are the upper and bottom edges of the IB.
The blue region represents the IB region.
The red dash-dotted lines are the Fermi level.}
\end{figure}

Figure \ref{fig:4}(a) shows that the Fermi level exists in the IB in the band
gap in whole the $x$ region, which suggests that the ferromagnetism is strongly
related to the IB.
At the MIT border ($x$=$\sim$2\%), the Fermi level is still in the band gap,
which suggests that the MIT occurs in the IB.
This behavior of the Fermi level is completely different from that in the case
of the heavily-doped $p$-type GaAs with the non-magnetic acceptors and
contradicts the VB conduction picture, where the IB completely merges into the
VB.
The IB still exists in the metallic GaMnAs, which means that the impurity states
 still remain in the band gap.
This is probably because the holes are trapped into the acceptor states induced
by $p$-$d$ hybridized orbitals after the Coulomb potential completely screened
\cite{mahadevan:2860, PhysRevLett.92.047201}.
At the onset of the ferromagnetism ($x$=$\sim$1\%), the $x$ dependence of the
Fermi level changes, which is thought to be induced by the change of the main
effect for determining the IB position from the screening effect to the
anti-crossing interaction.
This must give a clue to understanding the mechanism of the ferromagnetism.
We note that this anomalous behavior of the Fermi level is not directly related
to the MIT because the MIT and the turning up of the Fermi level position occur
at slightly different value of $x$ (1.5 - 2\% and $\sim$1\%, respectively).

In summary, we fabricated the DB heterostructures containing a
Ga$_{1-x}$Mn$_x$As QW with various $x$.
From the resonant tunneling measurements in these structures, we found that the
Fermi level exists in IB in the band gap in the whole range of $x$.
When $x$ is less than 1.0\%, the Fermi level becomes close to the VB top as $x$
increases, which is induced by the screening effect.
At $x$=1.0\% around the emergence of the ferromagnetism, the Fermi level becomes
closest to the VB, where $E_{\rm F}$ is 4 meV.
When $x$ is larger than 1.0\%, the Fermi level goes further away from the VB top
as $x$ increases, which is qualitatively explained by the anti-crossing
interaction.
The anomalous behavior of the Fermi level means that the ferromagnetic GaMnAs is
completely different from the non-magnetic heavily-doped $p$-type GaAs,
which indicates that the IB in the ferromagnetic GaMnAs is composed of the
$p$-$d$ hybridized orbitals.
Our results clearly show that the ferromagnetism and the MIT are strongly
related to the formation of the Mn-derived IB.

\begin{acknowledgments}
This work was partly supported by Grant-in-Aids for Scientific Research
including Specially Promoted Research, the Special Coordination Programs for
Promoting Science and Technology, and FIRST Program of JSPS.
\end{acknowledgments}

\bibliography{refs}

\end{document}